\documentclass{trbunofficial}
\usepackage{graphicx}
\usepackage{adjustbox}
\usepackage{subcaption}
\usepackage{algorithm}
\usepackage{amssymb}
\usepackage[noend]{algpseudocode}
\usepackage[thinc]{esdiff}
\usepackage[hidelinks]{hyperref}

\AuthorHeaders{Anagnostopoulos, Geroliminis}
\title{A Hybrid Microscopic Model for Multimodal Traffic with Empirical Observations from Aerial Footage}

\author{%
  \textbf{Georg Anagnostopoulos, Corresponding Author}\\
  Urban Transport Systems Laboratory (LUTS)\\
  EPFL, Lausanne, Switzerland, CH-1015\\
  georgios.anagnostopoulos@epfl.ch\\
  \hfill\break
  \textbf{Nikolas Geroliminis, Ph.D.}\\
  Urban Transport Systems Laboratory (LUTS)\\
  EPFL, Lausanne, Switzerland, CH-1015\\
  nikolas.geroliminis@epfl.ch
}

\begin{document}
\maketitle

\section{Abstract}
Microscopic traffic flow models can be distinguished in lane-based or lane-free depending on the degree of lane-discipline. This distinction holds true only if motorcycles are neglected in lane-based traffic. In cities, as opposed to highways, this is an oversimplification and it would be more accurate to speak of hybrid situations, where lane discipline can be made mode-dependent. Empirical evidence shows that cars follow the lanes as defined by the infrastructure, while motorcycles do not necessarily adhere to predefined norms and may participate in self-organized formation of virtual lanes. This phenomenon is the result of complex interactions between different traffic participants competing for limited space. In order to better understand the dynamics of modal interaction microscopically, we first analyze empirical data from detailed trajectories obtained by the pNEUMA experiment and observe patterns of mixed traffic. Then, we propose a hybrid model for multimodal vehicular traffic. The hybrid model is inspired by the pedestrian flow literature, featuring collision-free and anticipatory properties, and we demonstrate that it is able to reproduce empirical observations from aerial footage.

\hfill\break%
\noindent\textit{Keywords}: Anticipation, Collision-free, Drones, Mixed Traffic, Virtual Lane
\newpage

\section{Introduction}
Diversification of urban mobility leads to competition of different traffic modes for the limited road space as defined by the existing infrastructure. Empirical evidence from advanced sensors, coupled with appropriate traffic flow models can enhance our understanding of complex modal interactions that cannot be readily described by vehicular traffic theory. Revisiting the traditional traffic flow models can be also seen as the consequence of empirical evidence from new data collection methods, such as drone videography.

A fascinating application of drone videography in traffic is the pNEUMA dataset, which incorporates a massive collection of naturalistic vehicle trajectories captured by a swarm of drones in the center of Athens, Greece. Details of this experiment along with suggested applications are discussed in \cite{Barmpounakis2020a}. PNEUMA holds the promise of enabling new insights into multi-modal urban traffic, but at the same time, it also introduces a number of methodological challenges. Most notably, vehicle positions are not always matched to lanes. A more fine-grained organization of the data is therefore critical for a wide spectrum of downstream tasks, both on microscopic and macroscopic levels. We address this issue by formulating and solving a map matching problem and by introducing a novel methodology for detailed segmentation of vehicle trajectories based on steering events. As steering events, we define the critical points in time when a driver changes steering direction. Prerequisites for the determination of steering events are knowledge of the road network, matching of vehicles to road segments and information about kinematic characteristics, such as headings. Observed phenomena, such as lane formation, are then modeled microscopically.

Microscopic traffic flow models can be classified in two broad categories depending on the assumption of lane-discipline: lane-based \cite{Bando1995, Treiber2000, Newell2002, Kesting2007} and lane-free \cite{Kanagaraj2018, Papathanasopoulou2018, Sarkar2020, Papageorgiou2021}. This clear-cut distinction holds true only if motorcycles are neglected in the first case. In urban environments, as opposed to highways, this is an oversimplification and it would be more accurate to speak of hybrid situations of mode-dependent lane discipline. Motorcycles, or in general powered-two-wheelers (PTW), have unique kinematic characteristics \cite{Vlahogianni2014} and their interactions with cars are not yet well understood. For an extensive review of PTW literature, see \cite{Barmpounakis2016}.

In general, vehicular flow theory does not sufficiently cover the PTW case, because even the most fundamental traffic variables, such as density, are hydrodynamic in nature, whereas PTW has granular characteristics \cite{Maury2018}. A more adequate framework can be inspired by research in pedestrian flow, where experimental results \cite{Hoogendoorn2005,Seyfried2009, Zhang2011, Liao2014, Schadschneider2018} reveal striking resemblance to phenomena observed in PTW traffic and dedicated traffic variables have been developed \cite{Steffen2010,Liddle2011}. We distinguish mainly two kinds of microscopic pedestrian models that have been applied to PTW traffic: discrete choice and self-driven particle systems. Discrete choice models include the multinomial logit \cite{Lee2016, Gavriilidou2019, Sarkar2020} and the nested logit model \cite{Antonini2006, Robin2009, Shiomi2012}. In particular, \cite{Shiomi2012} investigates the case when flow is dominated by motorcycles, \cite{Lee2016} models a specific queuing scenario of motorcycles at an intersection based on lateral position, but without considering their orientation, and \cite{Gavriilidou2019} models a similar situation with bicycles. Self-driven particle systems, can be distinguished in first order models \cite{Tordeux2016, Xu2019, Zhang2021, Xu2021}, based on velocity (including heading), and second order models \cite{Helbing1995, Helbing2000}, based on acceleration. In motorized traffic, only second order models have been proposed \cite{Kanagaraj2018,Delpiano2020}. The potential of first order pedestrian models for modeling PTW movement remains unexplored. These models have several appealing properties, such as simple formulation, very few parameters, consideration of heading and are collision-free by construction. 

This paper is organized in two parts. The first section is dedicated to empirical observations and data segmentation. The last part introduces the hybrid model for mixed vehicular traffic flow. Our main focus is on the investigation, analysis and modeling of complex phenomena of self-organization, as exemplified by the formation of virtual lanes.

\section{Empirical observations and data segmentation}

In order to facilitate a thorough investigation of multi-modality in the pNEUMA dataset, given that it contains 25\% of motorcycles, our objective is to solve the trajectory segmentation problem for heterogeneous traffic, where lane discipline can be only partially assumed. In fact, it is reasonable to pose the problem in terms of self-organized lane formation instead of lane discipline. Starting from a macroscopic segmentation of all the vehicle trajectories, including motorbikes, we are interested to see, on a microscopic level, which parts of the trajectories are lane-keeping and which parts are devoted to maneuvering. Then, lanes in the most general sense of lane-keeping envelopes, can be easily identified as a direct consequence.

 In principle, vehicles never travel in perfectly straight lines and there is always some amount of curvature present. Lane-keeping requires regular corrections from the driver or rider who performs the steering. Surprisingly, the frequency of critical steering events drops drastically during the execution of lane-changing or other maneuvers. There is typically only one steering correction per maneuver. The existence of this sparsity is the main discovery of this paper and we will show that it can greatly simplify the challenging task of lane detection. This statement holds if we assume that the aforementioned critical events can be obtained. Of course, this is not generally possible or easy as it requires first, that vehicle headings are given, and second, that they are also devoid of noise, especially non-Gaussian anomalies. Very recent research on the pNEUMA dataset \cite{Kim2022}, discusses these matters in detail and we will assume for the reminder of the paper that perfect heading information is available.

\subsection{Macroscopic trajectory segmentation}
Because the pNEUMA dataset comes from an airborne experiment in a large urban area with more than 100 intersections, matching of vehicles to road segments augments the data with important contextual information such as road azimuth, traffic signal or bus stop locations.  Interestingly, in \cite{PAIPURI2021}, the authors postulate that even macroscopic analysis can benefit from map matching by facilitating the detection and exclusion of parked vehicles. It is known that parked vehicles are not considered as traffic participants in the sense of the two-fluid theory of town traffic \cite{Herman1979}.

The most successful method for macroscopic trajectory segmentation is the Hidden Markov Model (HMM). HMM \textit{"is an algorithm that can smoothly integrate noisy data and path constraints in a principled way"} \cite{Newson2009}. Initially, HMM was used for matching sparse GPS data, so, somewhat surprisingly, direct application of HMM map matching on dense trajectories from the pNEUMA dataset is problematic. This is due to the abundance of stationary observations, commonly referred to as stay points. In urban settings, stay points can be the result of low speeds during congestion, service related stops, the existence of traffic signals or emergency breaking due to random events. The static data entries, excluding parked vehicles, in the pNEUMA dataset are typically in the range of 40\% or more, which means that there is a great potential for compression. Consequently, we proceed as follows: each trajectory is divided in two parts: one static and one moving. This is a simple form of data reduction. The advantage of this approach is threefold. First, the computational burden on the HMM algorithm can be reduced.  Second, we can now make continuity assumptions for the moving part. Third, the method is lossless and the two trajectory partitions, static and moving, can be easily reconciled further downstream.

The HMM algorithm considers the set $N_r$ of road segments as a set of Markov states and vehicle positions as state measurements $z_t$, where the time $t$ is discrete. Given any noisy trajectory, the model recommends the most plausible sequence of state transitions. A solution is obtained within the constraints imposed by the connectivity of the underlying road network as provided by Open Street Maps (OSM) using an API \cite{Boeing2017}. We should stress that an optimized network configuration is key to the success of the matching. Because the pNEUMA dataset contains a large proportion of motorcycles, we choose the \verb"bike" preset with a few minor modifications.

\begin{figure}[!ht]
\centering
\begin{subfigure}{.675\textwidth}
    \centering
    \includegraphics[width=\textwidth]{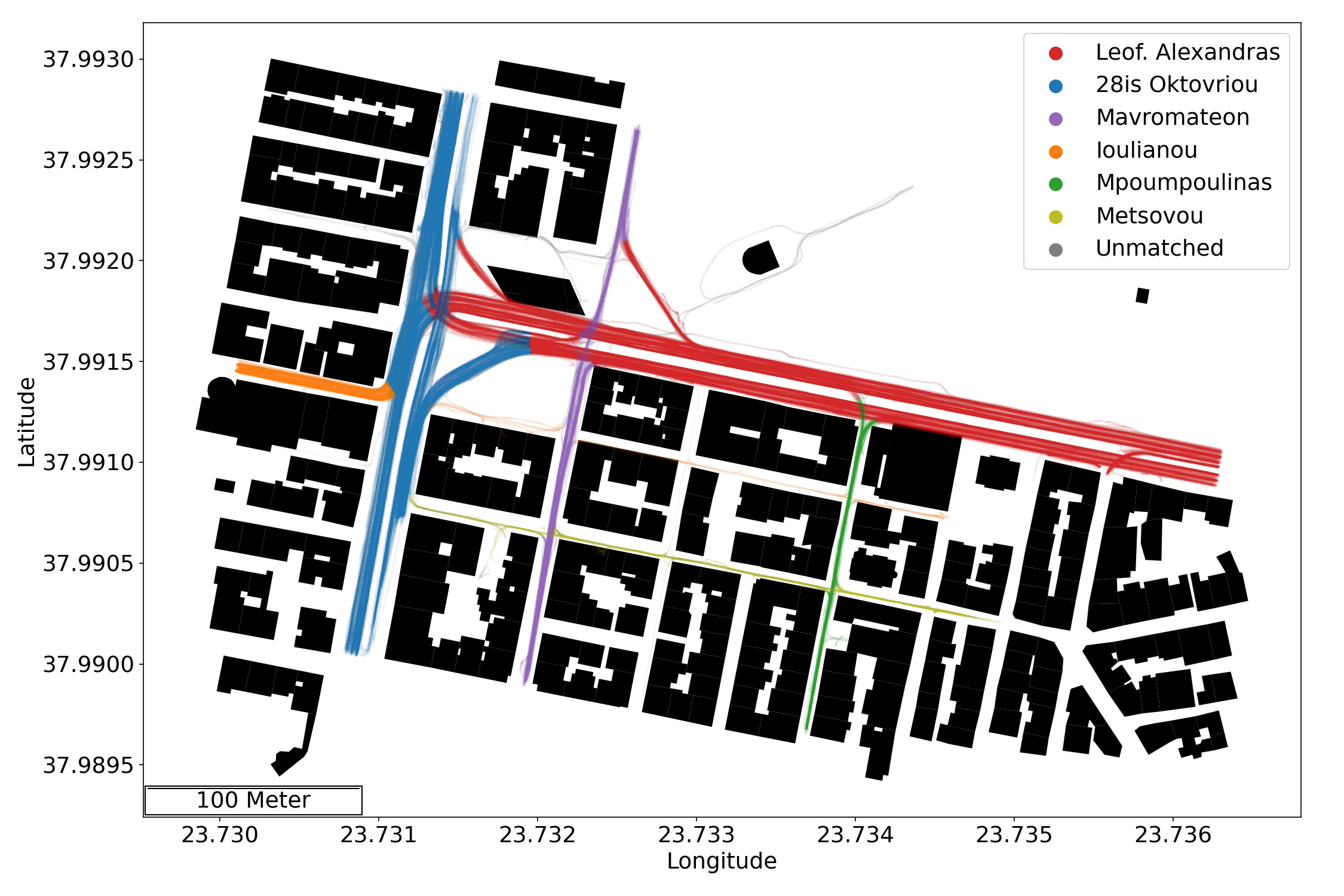}
    \caption{Street level segmentation.}\label{subfig:roads}
\end{subfigure}
\begin{subfigure}{.225\textwidth}
    \centering
    \includegraphics[width=\textwidth]{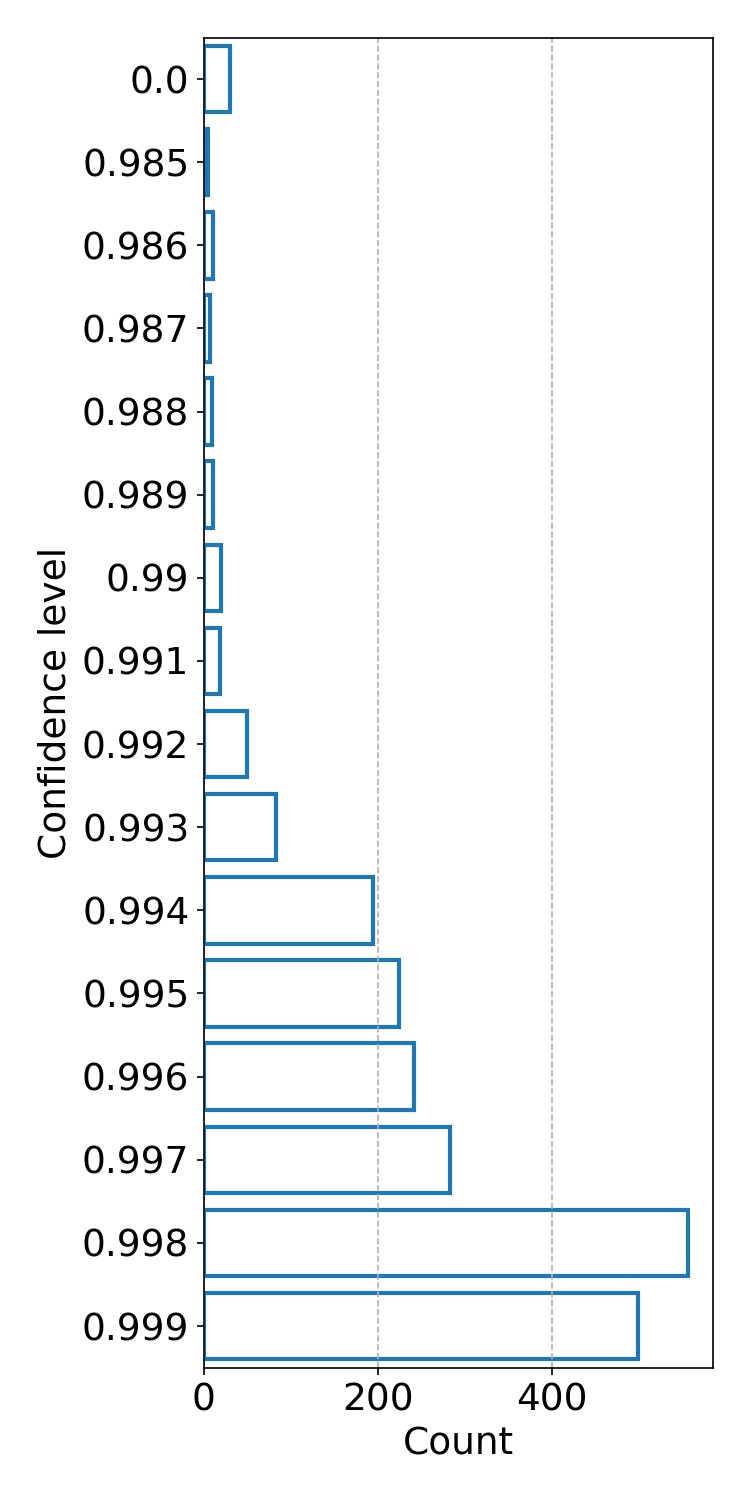}
    \caption{Confidence.}\label{subfig:histogram}
\end{subfigure}
\caption{Qualitative and quantitative assessment of the iterative HMM algorithm. }\label{fig:hmm}
\end{figure}

The most crucial aspect of HMM is the confidence level or minimum match probability. Setting this threshold universally for all trajectories is generally sub-optimal. Instead, we propose an iterative execution where we start with 100\% probability and reduce the confidence level in 0.1\% steps until the algorithm returns a valid sequence of states. If the probability drops below the 98.5\% threshold, as shown in Figure~\ref{subfig:histogram}, the trajectory is deemed unreliable and receives null confidence. A dynamic threshold tailored to each trajectory has significant performance advantages and is a good metric of matching trustworthiness. Finally, the output is a sequence of coordinates along different road segments. The primary direction of flow per link can be deduced from the median azimuth of the observations. Figure~\ref{subfig:roads} shows the result aggregated on the level of roads.

\subsection{Microscopic trajectory segmentation}
The maneuver detection method requires the following inputs: raw vehicle positions ${(x,y)}$, vehicle azimuth ${\theta_{vehicle}}$ and road azimuth ${\theta_{road}}$. Then, we can simply define the phase ${\phi=\theta_{vehicle}-\theta_{road}}$ and the one-dimensional gradient ${\nabla\phi = \diff{\phi}{s}}$ of the phase with respect to distance travelled ${s}$. The points with ${\nabla\phi = 0}$ are essentially inflection points, where the curvature changes sign or, in terms of driving/riding, a steering event occurs. This simple expression is inspired from the calculation of lateral acceleration, but without any temporal dependence. Our approach is therefore solely based on differential geometry.

\begin{figure}[!ht]
\centering
\begin{subfigure}{.9\textwidth}
    \centering
    \includegraphics[width=\textwidth]{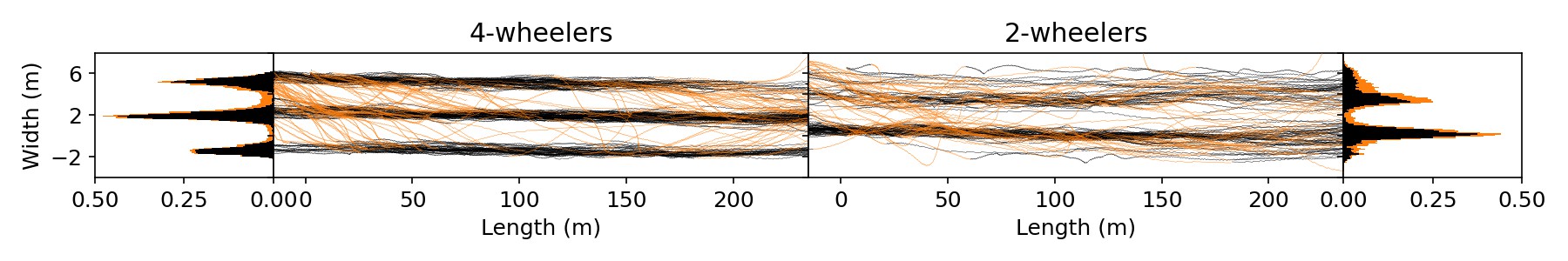}
    \caption{A 3-lane arterial, $250$m in length over a time period of $15$ minutes. Orange denotes maneuvers. Note that there is almost no spatial overlap between fixed and virtual lanes.}\label{subfig:lanes}
\end{subfigure}
\begin{subfigure}{.9\textwidth}
    \centering
    \includegraphics[width=\textwidth]{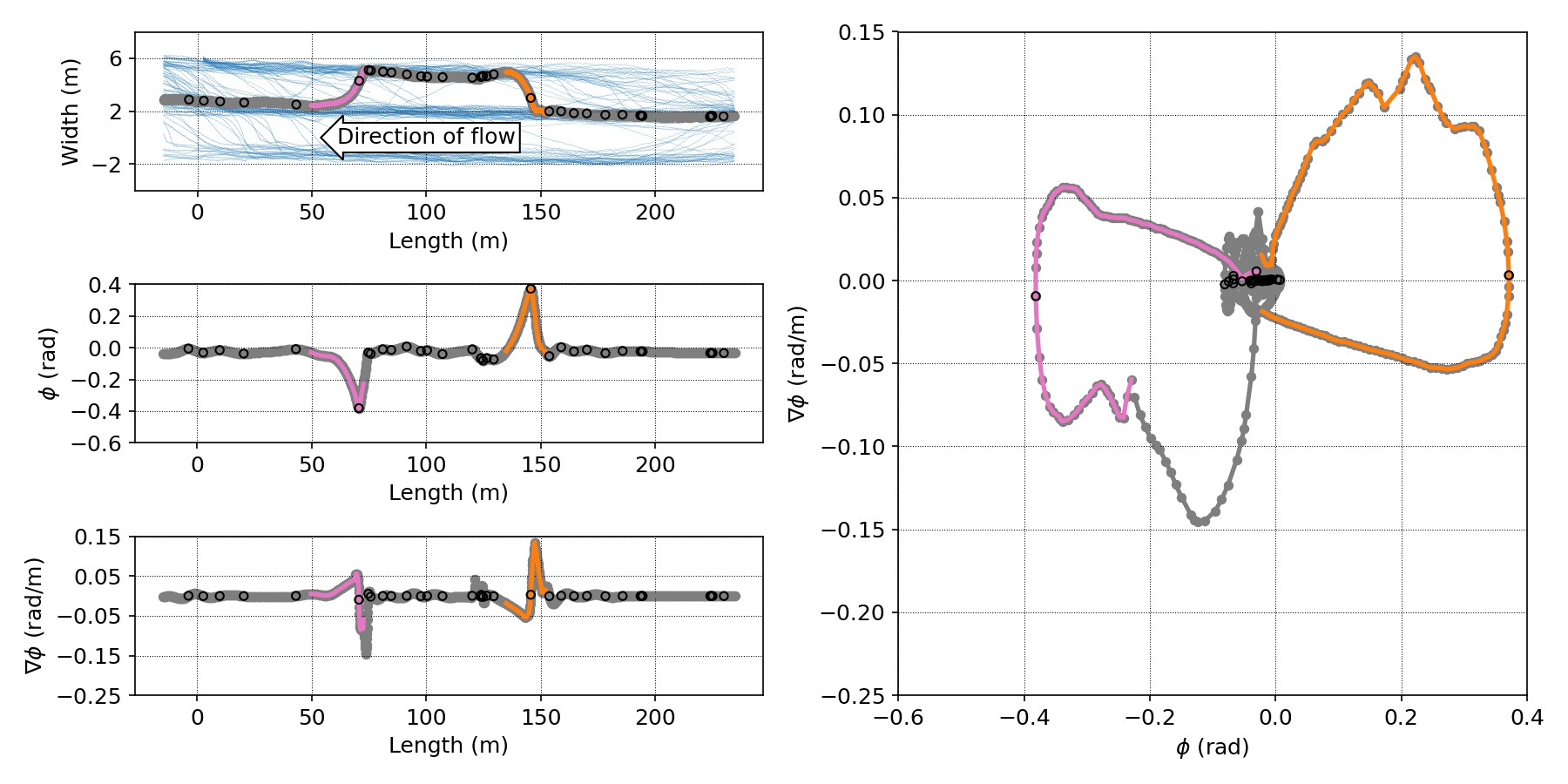}
    \caption{Illustration of the maneuver detection method based on sparse steering events. Regular events are concentrated near the origin and steering maneuvers form butterfly-like wings. The rightward maneuver is marked as orange and the leftward as pink.}\label{subfig:1315}
\end{subfigure}
\caption{Detailed trajectory segmentation for a 3-lane urban arterial. Lane discipline applies only to cars and other larger vehicles. Motorcycles do not follow the predefined lanes and may form emergent, virtual lanes in the available free spaces.}\label{fig:method}
\end{figure}

From empirical observations, it follows that between any consecutive pair of steering events, the trajectory can be either convex or concave. Then, convexity is determined directly from the raw data points by choosing the most central local extrema. If the central extremum is a minimum, the segment is convex and it is concave otherwise. The final step includes connecting the extrema and then checking for monotonic intervals. If the lateral shift exceeds $1$m, a maneuver is detected. This approach is devoid of systematic bias in $\phi$ and of temporal distortions introduced by the lateral acceleration, therefore it can detect a wide spectrum of maneuvers, Figure~\ref{fig:method}.

\begin{figure}[!ht]
\centering
\begin{subfigure}{.9\textwidth}
    \centering
    \includegraphics[width=\textwidth]{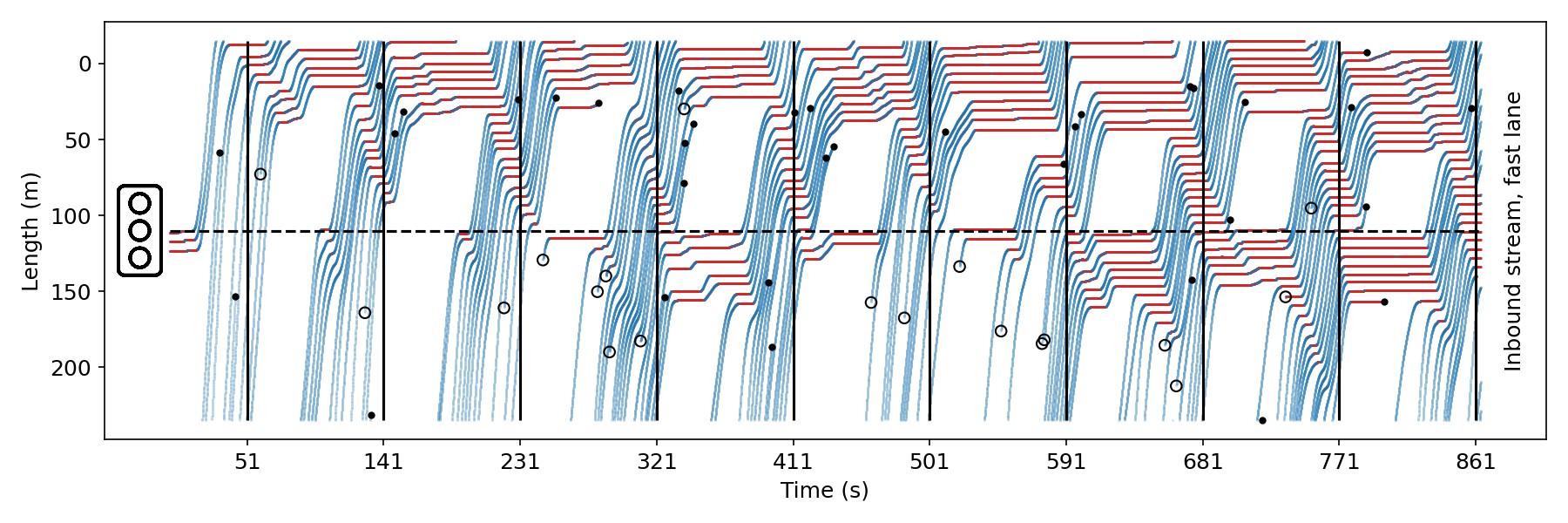}
    \caption{Time-space diagram for cars on the inbound fast lane, including lane-changing trajectories. Blue corresponds to moving cars and red to static cars. Open circle: vehicle enters the lane. Closed circle: vehicle exits the lane. The space axis follows convention.}\label{subfig:timespace}
\end{subfigure}
\begin{subfigure}{.9\textwidth}
    \centering
    \includegraphics[width=\textwidth]{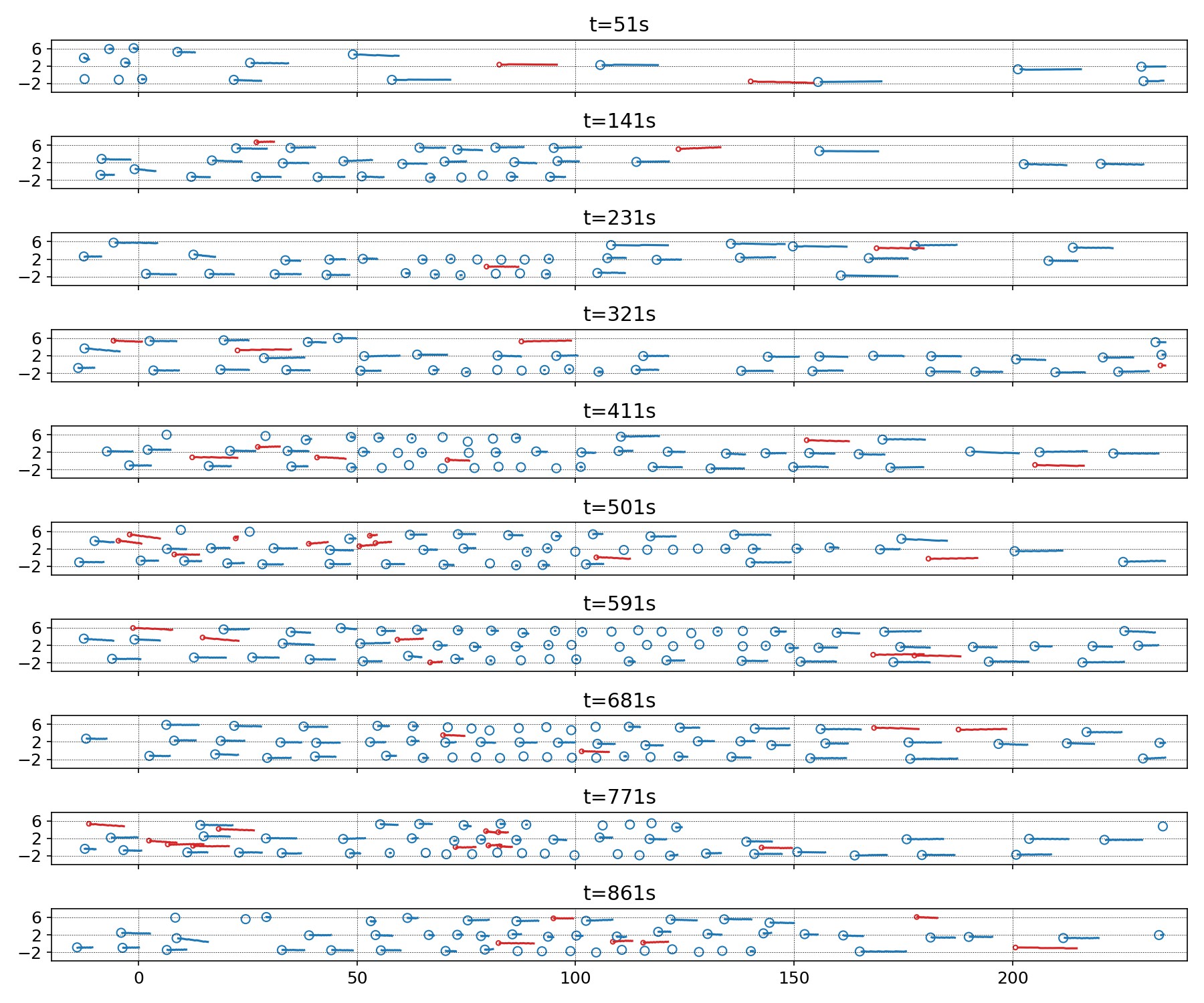}
    \caption{Periodic snapshots with minimal influence of traffic lights as obtained from the time-space diagram in Figure~\ref{subfig:timespace}. Velocity can be measured directly from the tails that extend 1s in the past. Blue corresponds to cars and red to motorcycles.}\label{subfig:snapshots}
\end{subfigure}
\caption{Generalized time-space diagram for cars and motorcycles. As congestion intensifies, motorcycles percolate forward by filtering through the free space between the standing cars.}\label{fig:ts}
\end{figure}

We distinguish between 2-wheelers (motorcycles) and an aggregation of the other modes (cars, taxis, medium/heavy vehicles and buses), jointly referred to as 4-wheelers. In the case of 4-wheelers, our method is able to accurately discern between 3 lane-keeping envelopes and sparse connections of these envelopes. Here we can clearly speak of lanes and lane-changes. This is true under ideal road conditions with straight roads of constant width. In the case of 2-wheelers, there are no fixed lanes, but virtual lanes can be formed under specific road and traffic conditions \cite{Vlahogianni2014}. The time-space diagram of Figure~\ref{subfig:timespace} shows no trace of artifacts, such as intersecting trajectories.

One advantage of this method, compared to \cite{Barmpounakis2020b}, is that the outcome is independent of the exact lateral boundaries around the lane-keeping envelope, as long as they do not cut into another lane. It is worth mentioning that our approach is parameter-free because it is not using clustering, smoothing or peak detection algorithms that have been previously proposed and it is thus based solely on the physical meaning of the driving or riding task. Additionally, a largely overlooked, but important detail is that even a very small drift between the empirical data and the theoretical routes on the map, may introduce systematic bias, leading to significant under- or overestimation. Our method is designed to overcome this problem.

\begin{figure}[!ht]
\centering
\begin{subfigure}{.45\textwidth}
    \centering
    \includegraphics[width=\textwidth]{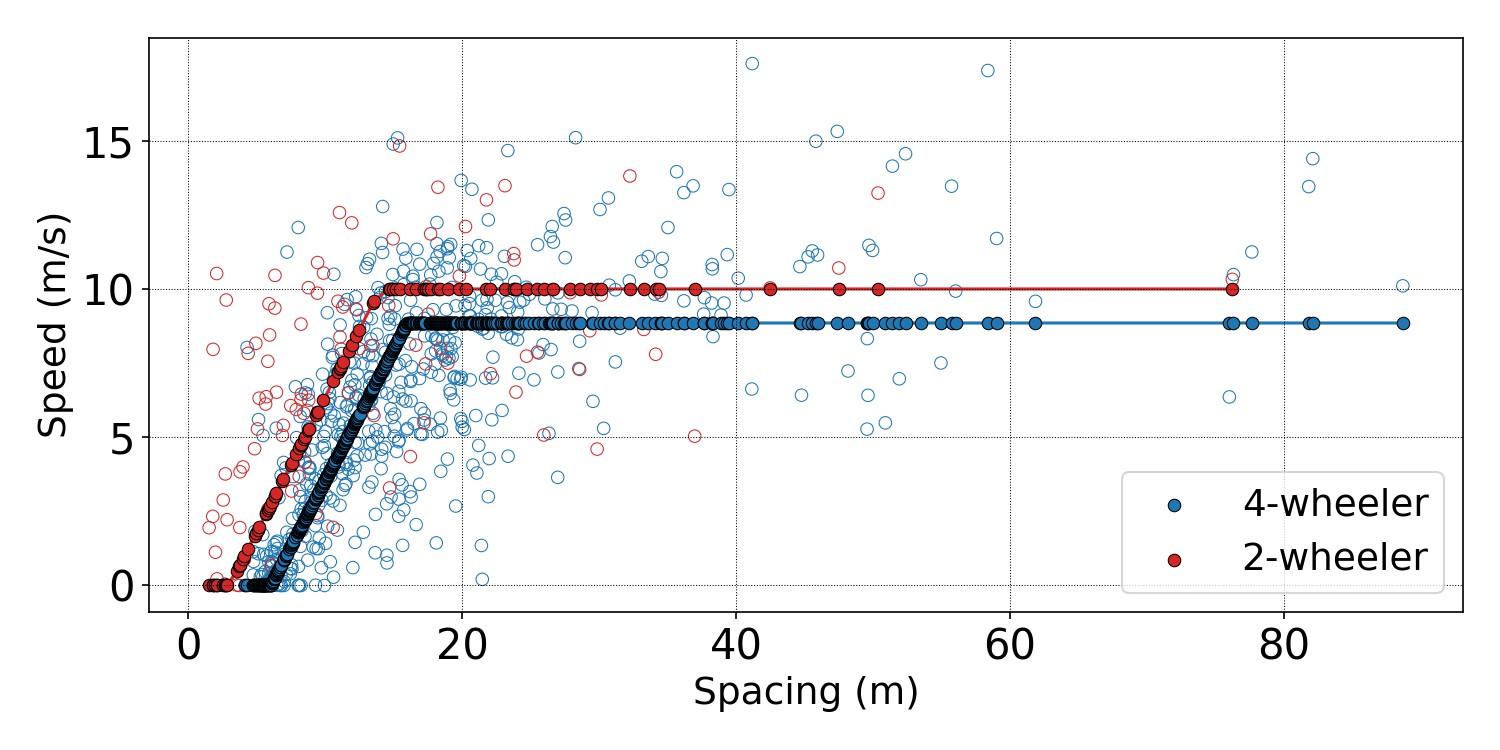}
    \caption{The fundamental relationship between speed and spacing as obtained from the snapshots in Figure~\ref{subfig:snapshots}. Piecewise linear regression.}\label{subfig:fd}
\end{subfigure}\hspace{5mm}
\begin{subfigure}{.45\textwidth}
    \centering
    \adjincludegraphics[width=\textwidth,trim={500 410 500 200},clip]{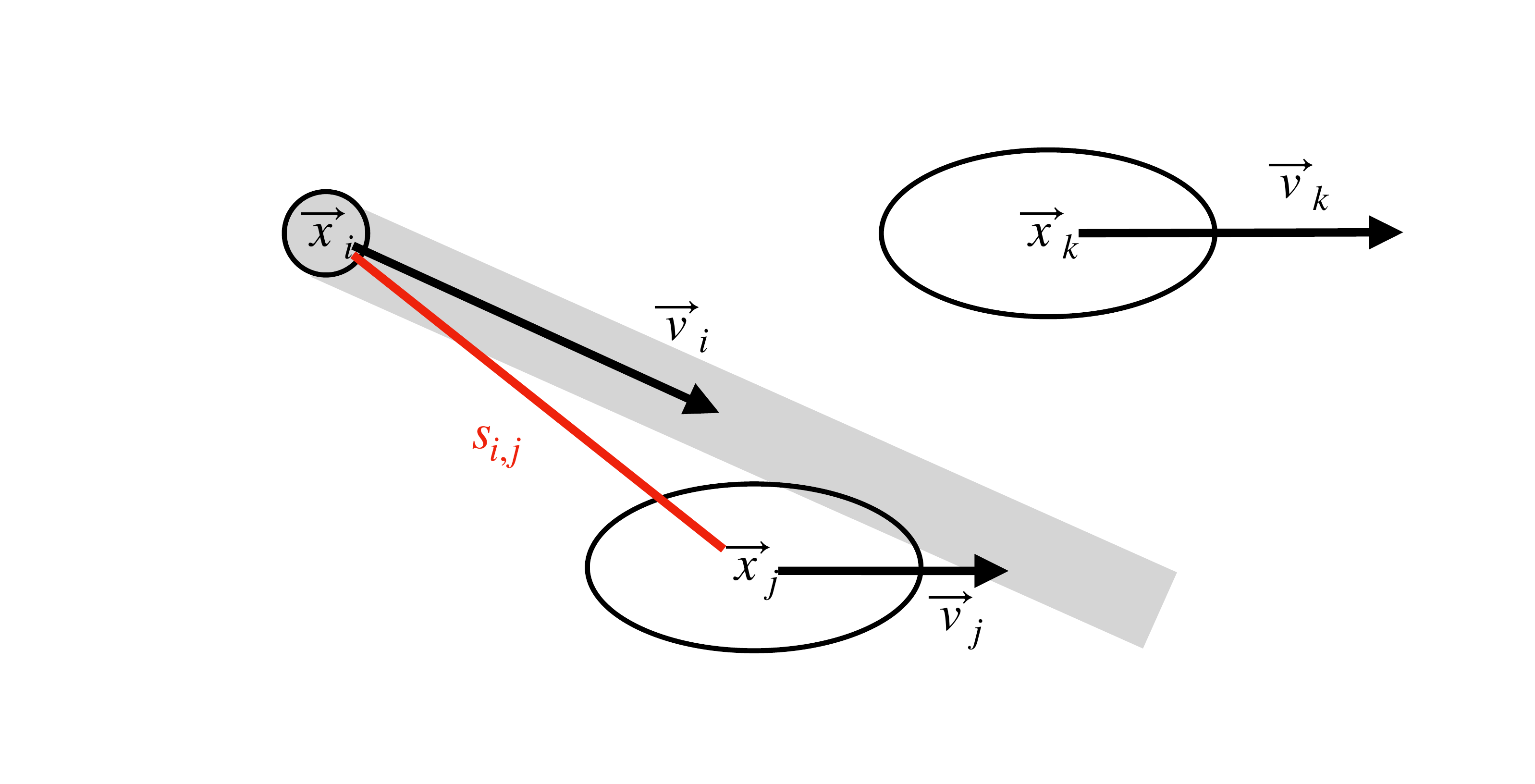}
    \caption{Principle for estimation of the spacing $s_{i,j}$. We are considering the center-to-center distance to the closest longitudinally overlapping vehicle.}\label{subfig:spacing}
\end{subfigure}
\begin{subfigure}{.45\textwidth}
    \centering
    \includegraphics[width=\textwidth]{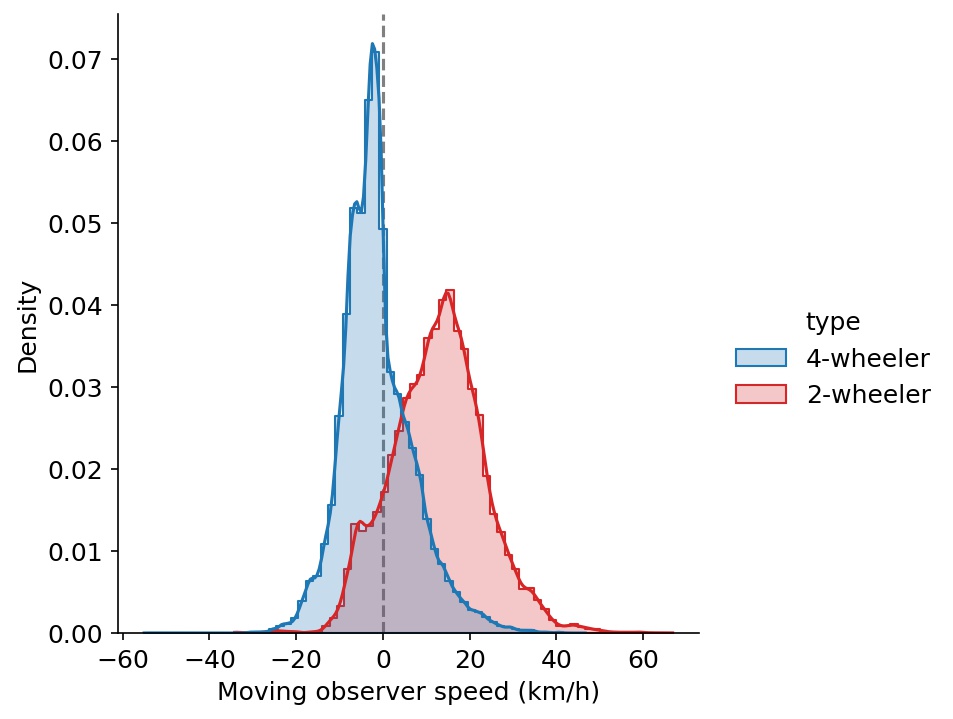}
    \caption{PDF of the moving observer speed in a given spatiotemporal neighborhood.}\label{subfig:tube_pdf}
\end{subfigure}\hspace{5mm}
\begin{subfigure}{.45\textwidth}
    \centering
    \includegraphics[width=\textwidth]{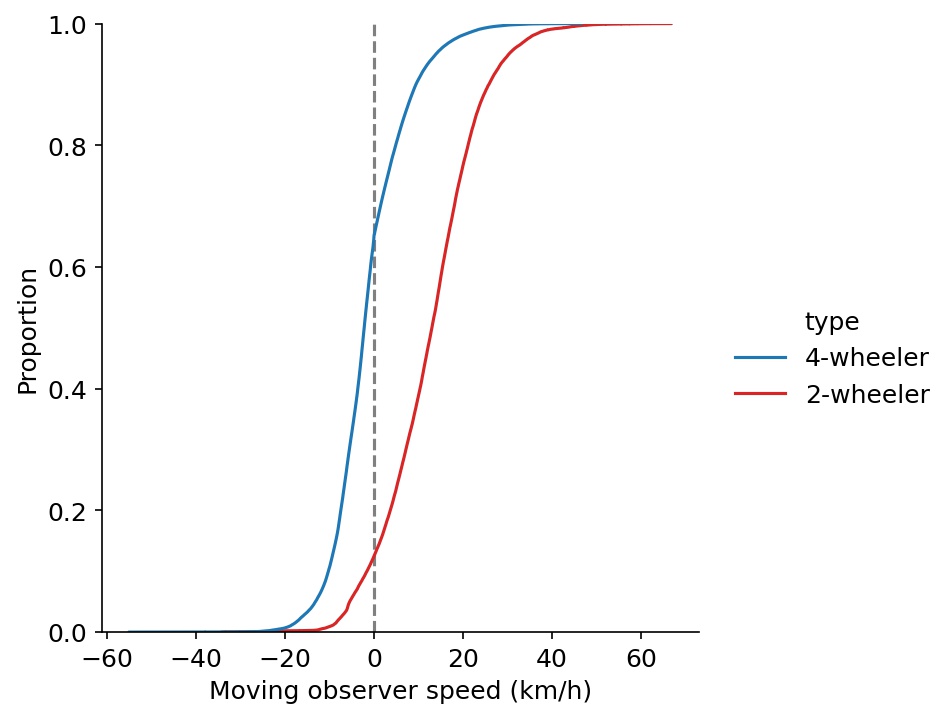}
    \caption{The CDF suggests that 70\% of the cars are slower than their spatiotemporal neighbours.}\label{subfig:tube_cdf}
\end{subfigure}
\caption{Multimodal microscopic fundamental relationship and multimodal speed dynamics.}\label{fig:data}
\end{figure}

For the hybrid model, that is introduced in the following section, we will estimate some parameters from the data. As a first approximation, we assume that the effect of traffic signals can be neglected. In order to neutralize their influence, spatial snapshots are taken at appropriate time intervals, as shown in Figure~\ref{subfig:snapshots}, which is a generalized time-space diagram that can also capture lane-free traffic. The fixed signal settings have been extracted by \cite{EspadalerClapes2022} and give a signal cycle of $90$s. We estimate the fundamental relationship between speed and spacing \cite{Newell2002} from the aforementioned snapshots, using the spacing definition from Figure~\ref{subfig:spacing}. A piece-wise linear fit is applied \cite{Roelfs2022}. A hybrid modeling approach is further motivated by speed dynamics. This becomes clear if we consider the moving observer speed in a given spatiotemporal neighborhood, here heuristically defined as $60\text{m}\times2\text{s}$. Figure~\ref{subfig:tube_cdf} reveals ashtonishing differences in speed between the two modes.  

\section{Towards a hybrid model}
From the previous analysis it is clear that lane discipline is not only a matter of culture, but it can simply be mode-dependent. Larger modes, such as cars, buses and trucks, follow lanes that are predefined by the infrastructure. On the other hand, PTW participate in a lane formation process that is self-organized, meaning that the virtual lanes are emerging as a result of some physical mechanism, rather than compliance to an agreed-upon social norm \cite{Schadschneider2018}. While numerical simulations of self-organization in pedestrian counter-flows have a history of at least 30 years \cite{Helbing1995}, in vehicular traffic, virtual lanes and the related percolation phenomena of creeping or filtering have been studied mostly statistically in terms of their possible determinants \cite{Vlahogianni2014} or on a macroscopic level by a small branch of research on the so-called porous model \cite{Nair2011,Ambarwati2014}. The dynamics of lane formation are rarely discussed in the context of vehicular traffic. To fill this gap, we propose a hybrid model for mixed vehicular traffic that is inspired by pedestrian dynamics.

\subsection{Experimental setup and initial conditions}
\begin{figure}[!ht]
  \centering
  \adjincludegraphics[width=0.9\textwidth,trim={0 120 0 120},clip]{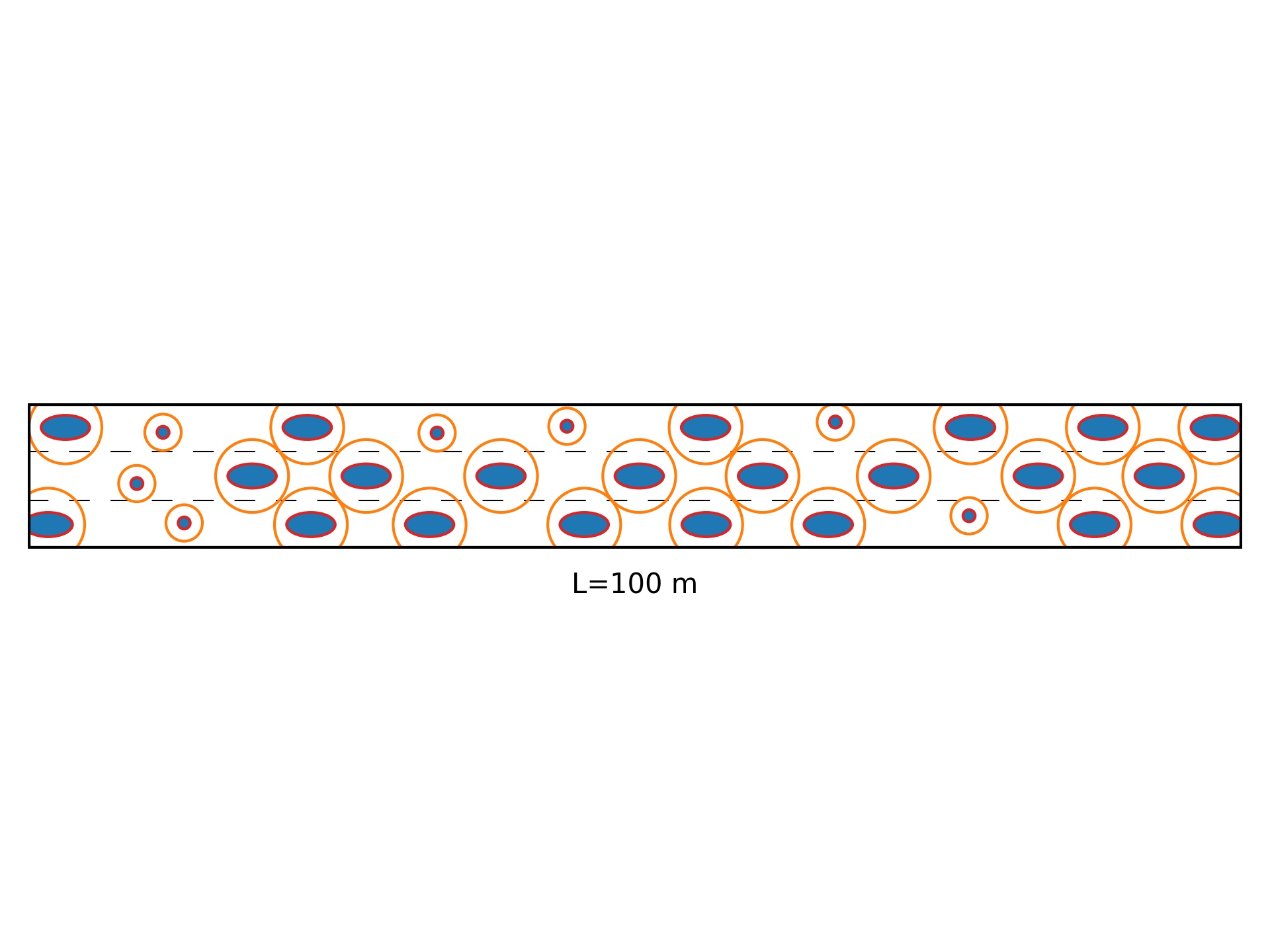}
  \caption{Blue noise initialization by the Poisson disk sampling algorithm. Radii are chosen from a discrete set with a given probability. Larger vehicles are laterally constrained to follow lanes of 4m. Motorcycles are placed randomly in both dimensions and are allowed to move freely.}\label{fig:blue}
\end{figure}

We consider a simple experimental setup of a straight road segment 100m in length with 3 lanes of 4m width each. We assume periodic boundary conditions and congested traffic conditions with a density of 100veh/km/lane. The road is used by two different modes: cars and motorcycles. Cars are modeled as $2\times4$m ellipses and the motorcycles are modeled as circles of 1m in diameter. Perfect lane discipline is assumed for the cars and any lane-changes are neglected. On the other hand, motorcycles are allowed to move freely in two dimensions, see Figure~\ref{fig:blue}.

The initial positions are generated by stochastic sampling, also known as dart-throwing or blue noise. The Poisson disk sampling algorithm due to \cite{Bridson2007}, that is widely used in the computer games industry, is the established method for efficient dart-throwing. For a given domain, in our case a subset of $\mathbb{R}^{2}$, a background lookup grid is used for accelerated neighborhood calculations. The minimum distance between samples $r$, which is constant in the original method, determines the cell size of the substrate. A cell size of $r/\sqrt{2}$ guarantees that each cell may contain at most one sample. There exist variable radii adaptations of the Poisson disk sampling algorithm \cite{Tulleken2008,Dwork2021}, where the minimum radius is a function of the spatial domain. Our approach is based on the fixed radius implementation due to \cite{Hill2020}, but we are drawing radii randomly from a discrete set $\{r_{min}=3m,r_{max}=6m\}$, with given probabilities $\{0.25,0.75\}$. Please note that when multiple radii are used, the grid cell is determined by the largest radius and each cell is allowed to contain a list of samples \cite{Tulleken2008}. In our case, the largest radii also receive lateral constraints for lane alignment. 

\subsection{Anticipation velocity model for mixed vehicular traffic}
The hybrid model builds on the anticipation velocity model (AVM) \cite{Xu2021}, a very recently proposed pedestrian model with good lane formation capabilities. The AVM belongs to a family of collision-free first order models based on velocity \cite{Tordeux2016, Xu2019, Zhang2021, Xu2021}, which follow the mathematical framework laid out by the authors of \cite{Maury2018}. In contrast to the AVM, where pedestrian agents are represented by disks of constant radius, the hybrid model distinguishes between two classes of vehicular agents: cars and motorcycles. Cars have elliptical shape and move in predefined lanes. Motorcycles are represented as disks and their movement is two-dimensional. For each agent $i$, we define the position $\vec{x_i}$ and the velocity $\vec{v_i}$, such that $\vec{v_i}=\dot{\vec{x_i}}$. This is an ordinary differential equation (ODE) that can be solved if we write $\vec{v_i}=\vec{e_i}\cdot v_i$, where $\vec{e_i}$ is the heading and $v_i$ is the speed of agent $i$, calculate each of them and then integrate. A navigation module captures the dynamics of $\vec{e_i}$ and a speed module the evolution of speed $v_i$. The Euler integration scheme will be used for simulations.

For the navigation, we define a target direction $\vec{e_i}^0(t)$, which in this case tends to maintain the initial lateral position, and a set $N$ of interacting agents visually perceived by the driver/rider:

\begin{linenomath}
  \begin{equation} \label{eq:1}
  N_i(t) = \left\{ j,\;\left(\text{condition}^1\right)\;\text{or}\;\left(\text{condition}^2\right)\;\text{or}\;\left(\text{condition}^3\right)\;\text{or}\;\left(\text{condition}^4\right) \right\},
  \end{equation}
\end{linenomath}
where $\text{condition}^1={\vec{e_i}(t)\cdot\vec{e}_{i,j}(t)> 0}$ captures interactions in the current direction of movement, $\text{condition}^2={\vec{e_{i}}^0(t)\cdot\vec{e}_{i,j}(t)> 0}$ accounts for interactions in the intended direction of movement,\newline $\text{condition}^3=\left\vert{\vec{e_i}(t)\cdot\vec{e}_{i,j}(t)}\right\vert\leq \frac{r_i+r_j}{s_{i,j}(t)}$ and $\text{condition}^4=\left\vert{\vec{e_i}^0(t)\cdot\vec{e}_{i,j}(t)}\right\vert\leq \frac{r_i+r_j}{s_{i,j}(t)}$ are active in case of laterally overlapping agents. Here $s_{i,j}(t)$ denotes the spacing between agents $i$ and $j$ and $\vec{e}_{i,j}(t)$ is the unit vector from $j$ to $i$. Note that $r_i,r_j$ are the elliptical radii in the direction of the respective interaction. For better readability, we will not write the parametric equation of $r$ explicitly.

In the reminder of the navigation equations we are assuming that agent $i$ is a motorcycle, but agent $j$ can also be a car. After establishing a perception of the actual situation, the rider attempts to forecast the near future before making a decision. This anticipation strategy contributes considerably to the realism of the model. The anticipated spacing after a prediction horizon $t^a$ is

\begin{linenomath}
  \begin{equation} \label{eq:2}
  s_{i,j}^a(t+t^a) = \max\left\{ \left(r_i+r_j\right),\;\left(\vec{x_{j}}^a(t+t^a)-\vec{x_{i}}^a(t+t^a)\right)\cdot\vec{e}_{i,j}(t)\right\},
  \end{equation}
\end{linenomath}
where $\vec{x_{i}}^a(t+t^a)=\vec{x_{i}}(t)+\vec{v_{i}}(t)\cdot t^a$ denotes the predicted position. The difference of predicted positions is projected on the vector of interaction $\vec{e}_{i,j}(t)$. The anticipated gap cannot be less than the sum of the radii, in which case there would be a collision.

If we assume that the agents are in physical contact, then $\left(r_i+r_j\right)=s_{i,j}^a(t+t^a)$. This situation corresponds to minimum comfort, which in turn can be translated as maximum repulsion. The dimensionless repulsion $R(t)$ drops exponentially with increasing spacing and depends on a range parameter $D$, as well as an anisotropic intensity function $a(t)$.

\begin{linenomath}
  \begin{equation} \label{eq:3}
  R_{i,j}(t) = a_{i,j}(t)\cdot\exp{\left(\frac{\left(r_i+r_j\right)-s_{i,j}^a(t+t^a)}{D}\right)},\;D>0,
  \end{equation}
\end{linenomath}

The function $a_{i,j}(t)$ is equal to a positive intensity parameter $k$ when $\vec{e_{i}}^0(t)\cdot\vec{e}_j(t)=1$, and it is equal to $2k$ when $\vec{e_{i}}^0(t)\cdot\vec{e}_j(t)=-1$. This anisotropy captures the severity of a head-on collision.
\begin{linenomath}
  \begin{equation} \label{eq:4}
  a_{i,j}(t) = k\left(1 + \frac{1-\vec{e_{i}}^0(t)\cdot\vec{e}_j(t)}{2}\right),\;k>0,
  \end{equation}
\end{linenomath}

In order to exclude the possibility of backwards movement, the repulsion acts on the normal vector $\vec{n}_{i,j}(t) $, that is parallel to $\vec{e_i}^{0\bot}(t)$, such that $\vec{e_i}^{0\bot}(t)\cdot \vec{e_{i}}^0(t)=0$. It is specified as

\begin{linenomath}
  \begin{equation} \label{eq:5}
  \vec{n}_{i,j}(t) = -\text{sign}\left(\vec{e_{i,j}}^a(t+t^a)\cdot\vec{e_i}^{0\bot}(t)\right)\cdot\vec{e_i}^{0\bot}(t),
  \end{equation}
\end{linenomath}
where $\vec{e_{i,j}}^a(t+t^a)=\vec{x_{j}}^a(t+t^a)-\vec{x_{i}}(t)$ is a vector emanating from the center of $i$ to the predicted position of agent $j$. Equation \eqref{eq:5} means that the sense of the repulsion depends on the future and not the current position of the repeller.

The influence of a road curb $w$ can be modeled in a similar way \cite{Xu2019}, but the corresponding equations simplify considerably because the obstacle is static and anticipation is not applicable. Interactions with the road boundaries will be penalized with $2k$, thus there is also no anisotropy.

\begin{linenomath}
  \begin{equation} \label{eq:6}
  R_{i,w}(t) = k_w\cdot\exp{\left(\frac{r_i-s_{i,w}(t)}{D_w}\right)},\;k_w=2k>0,\;D_w=D>0,
  \end{equation}
\end{linenomath}

\begin{linenomath}
  \begin{equation} \label{eq:7}
  \vec{n}_{i,w}(t) = -\text{sign}\left(\vec{e_{i,w}}(t)\cdot\vec{e_i}^{0\bot}(t)\right)\cdot\vec{e_i}^{0\bot}(t),
  \end{equation}
\end{linenomath}
where $\vec{e_{i,w}}(t)=C_{w}(t)-\vec{x_{i}}(t)$, $C_w(t)$ is the closest point on the curb,  $\vec{e_{i,w}}(t)$ is the unit vector defined by $\vec{x_{i}}(t),\;C_{w}(t)$ and $(r_i-s_{i,w}(t))$ is the distance to the curb. From the equations \eqref{eq:3}, \eqref{eq:5}, \eqref{eq:6} and \eqref{eq:7} the desired direction of movement $\vec{e_{i}}^d(t)$ can be calculated as the weighted average of the intended direction (destination) and the lateral deviations due to repulsion from other agents or road edges:

\begin{linenomath}
  \begin{equation} \label{eq:8}
  \vec{e_{i}}^d(t) = u\left(\vec{e_{i}}^0(t) + \sum_{j \in N_i(t)}R_{i,j}(t)\cdot\vec{n}_{i,j}(t) + \sum_{w \in W}R_{i,w}(t)\cdot\vec{n}_{i,w}(t)\right),
  \end{equation}
\end{linenomath}
where $u$ is a normalization constant such that $\|\vec{e_{i}}^d(t)\|=1$ and $W$ is the set containing the road curbs. The navigation module is concluded by calculating the rate of change of the heading as a negotiation between the current and the desired direction, considering also a relaxation term $\tau$.

\begin{linenomath}
  \begin{equation} \label{eq:9}
  \diff{\vec{e_{i}}(t)}{t} = \frac{\vec{e_{i}}^d(t)-\vec{e_{i}}(t)}{\tau}.
  \end{equation}
\end{linenomath}

Finally, the speed module is responsible for the longitudinal dynamics \cite{Xu2019,Xu2021}. The set of imminently colliding agents, for both motorcycles and cars, is based on the updated direction $\vec{e_{i}}(t)$:
\begin{linenomath}
  \begin{equation} \label{eq:10}
  J_i(t) = \left\{ j,\;{\vec{e_i}(t)\cdot\vec{e}_{i,j}(t)\geq 0}\;\text{and} \left\vert{\vec{e_i}^{\bot}(t)\cdot\vec{e}_{i,j}(t)}\right\vert\leq \frac{r_i+r_j}{s_{i,j}(t)}\right\},
  \end{equation}
\end{linenomath}
where the first condition is similar to $\left(\text{condition}^1\right)\;$ and the second condition captures longitudinally overlapping agents (leaders). Note that in the case of motorcycles, we also need to take into account the set of imminently colliding road curbs. This is not necessary for cars as they move straight.

\begin{linenomath}
  \begin{equation} \label{eq:11}
  JW_i(t) = \left\{ w,\;{\vec{e_i}(t)\cdot\vec{e}_{i,w}(t)\geq 0}\right\},
  \end{equation}
\end{linenomath}

Then the maximum collision-free distance in the updated direction of movement can be calculated by Eq. \eqref{eq:12} for the set $J$ and by Eq. \eqref{eq:13} for the set $JW$, where $a_w$ is the angle to curb:

\begin{linenomath}
  \begin{equation} \label{eq:12}
  s_i(t) = \min_{j \in J_i(t)} s_{i,j}(t)-(r_i+r_j),
  \end{equation}
\end{linenomath}

\begin{linenomath}
  \begin{equation} \label{eq:13}
  sw_i(t) = \min_{w \in JW_i(t)} \frac{s_{i,w}(t)-r_i}{\cos{a_w}}.
  \end{equation}
\end{linenomath}

The speed is updated by the Optimal Velocity Model (OVM) \cite{Bando1995}, which is used here in the first order \cite{Tordeux2016}. This is one of the simplest models based on the speed-spacing relationship.

\begin{linenomath}
  \begin{equation} \label{eq:14}
  v_i(t) = \min\left\{ v_i^0,\;{\max\left\{\epsilon,\frac{s_i(t)}{T}\right\}},\;{\max\left\{\epsilon,\frac{sw_i(t)}{T}\right\}}\right\},
  \end{equation}
\end{linenomath}

where $v_i^0$ is the desired speed, $\epsilon$ is the machine epsilon, $T$ is the time gap and $s_i(t),\;sw_i(t)$ follow from Eqs. \eqref{eq:12} and \eqref{eq:13}. Using the machine epsilon instead of $0$ facilitates implementation because the velocity vector will not vanish completely, even during a full stop. To summarize, several features not present in the AVM have been incorporated in the hybrid model, such as multimodality, ellipses instead of circles, interactions with laterally overlapping agents and the influence of the road curbs. 

\subsection{Recommended parameters for numerical simulations}
The parameters that are adopted for numerical simulations with the hybrid model are summarized in Table~\ref{tab:versions} along with some reference values from the AVM model. The integration time step $\Delta t$, as well as the relaxation parameter $\tau$ remain unchanged. Simulations with the hybrid model have confirmed that $\tau = 0.3$s gives smooth results without sacrificing the stability of the solution that must remain collision-free. As we mentioned in the description of our experimental setup, instead of a fixed radius $r$, the hybrid model features ellipses that can vary in their semi-axes. Given that we are considering periodic boundary conditions, this can be a problem when vehicles exit the system at an arbitrary incidence angle. We therefore model motorcycles as disks of $r=0.5$m. Cars are modeled as ellipses because the incidence angle is always the same and the difference in dimensions cannot be neglected: accurate spacing is indispensable for realistic queue propagation.

\begin{table}[!ht]
	\caption{Comparison of model parameters for numerical simulations.}\label{tab:versions}
	\begin{center}
		\begin{tabular}{l l l l l l l l l}
			Model & ${r}$ [m] & ${k}$  & ${D}$ [m] & ${T}$ [s] & ${\Delta t}$ [s] & ${\tau}$ [s] & ${t^{a}}$ [s] & ${v_0}$ [m/s]\\\hline
			AVM   & \begin{tabular}{c}0.18\end{tabular} & 3.0 & 0.1 & 1.06 & 0.05 & 0.3 & 1.00 & \begin{tabular}{c}${N \sim (1.55,0.18^2)}$\end{tabular} \\
			Hybrid   & \begin{tabular}{c}0.5-0.5 \\1.0-2.0\end{tabular} & 0.1 & 3.0 & 0.90 & 0.05 & 0.3 & 0.88 &  \begin{tabular}{c}10.0\\8.85\end{tabular} \\\hline
		\end{tabular}
	\end{center}
\end{table}

A significant difference in shape, dimension and speed between pedestrian and vehicular agents, also signifies a change by one order of magnitude in the repulsion parameters. The range of interactions is much larger for vehicles and simulation shows satisfactory results when $D=3$m. On the other hand, the intensity of the repulsion is much lower for vehicular agents, where pushing phenomena, as observed in pedestrian crowds, simply do not exist. The rest of the parameters can be estimated directly from the data. We observe that the slope of the speed-spacing diagram in Figure~\ref{subfig:fd}, known as time gap, $T=0.9$s is lower than the values for town traffic given in \cite{Treiber2013}. Please note that $T$ also gives an upper bound for the anticipation horizon $t^a<T$, a condition that is met in both models. Furthermore, the data indicate a higher free-flow speed for motorcycles. Because cars are not allowed to overtake in our model, we refrain from using stochastic terms for the desired velocity. Lane-changing features for cars will be added in our future work. Finally, the fit from Figure~\ref{subfig:fd} also shows that the critical spacing is smaller for motorcycles than for cars.

\section{Results and discussion}
\begin{figure}[!ht]
\centering
\begin{subfigure}{.45\textwidth}
    \centering
    \includegraphics[width=\textwidth]{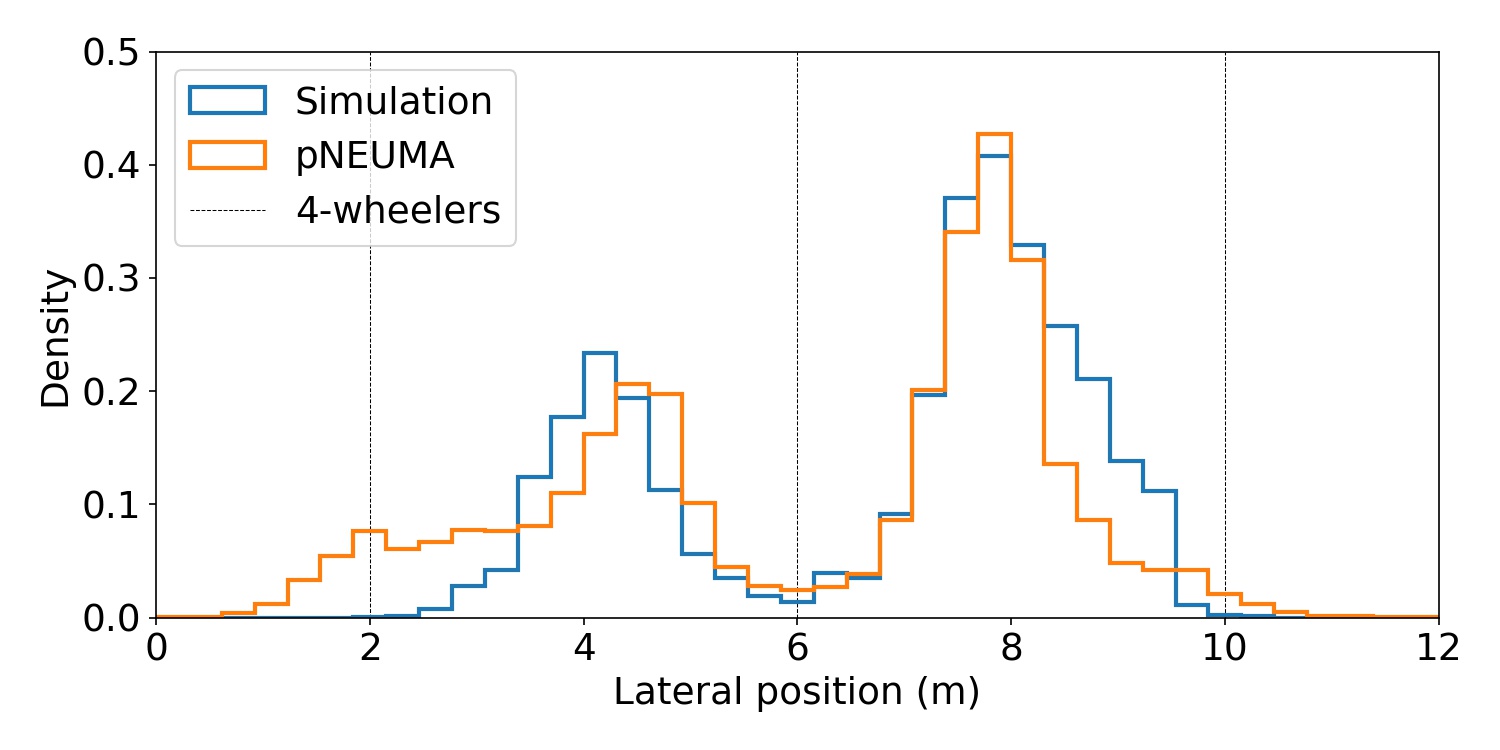}
    \caption{Lane formation in data and simulation.}\label{subfig:step}
\end{subfigure}
\begin{subfigure}{.45\textwidth}
    \centering
    \includegraphics[width=\textwidth]{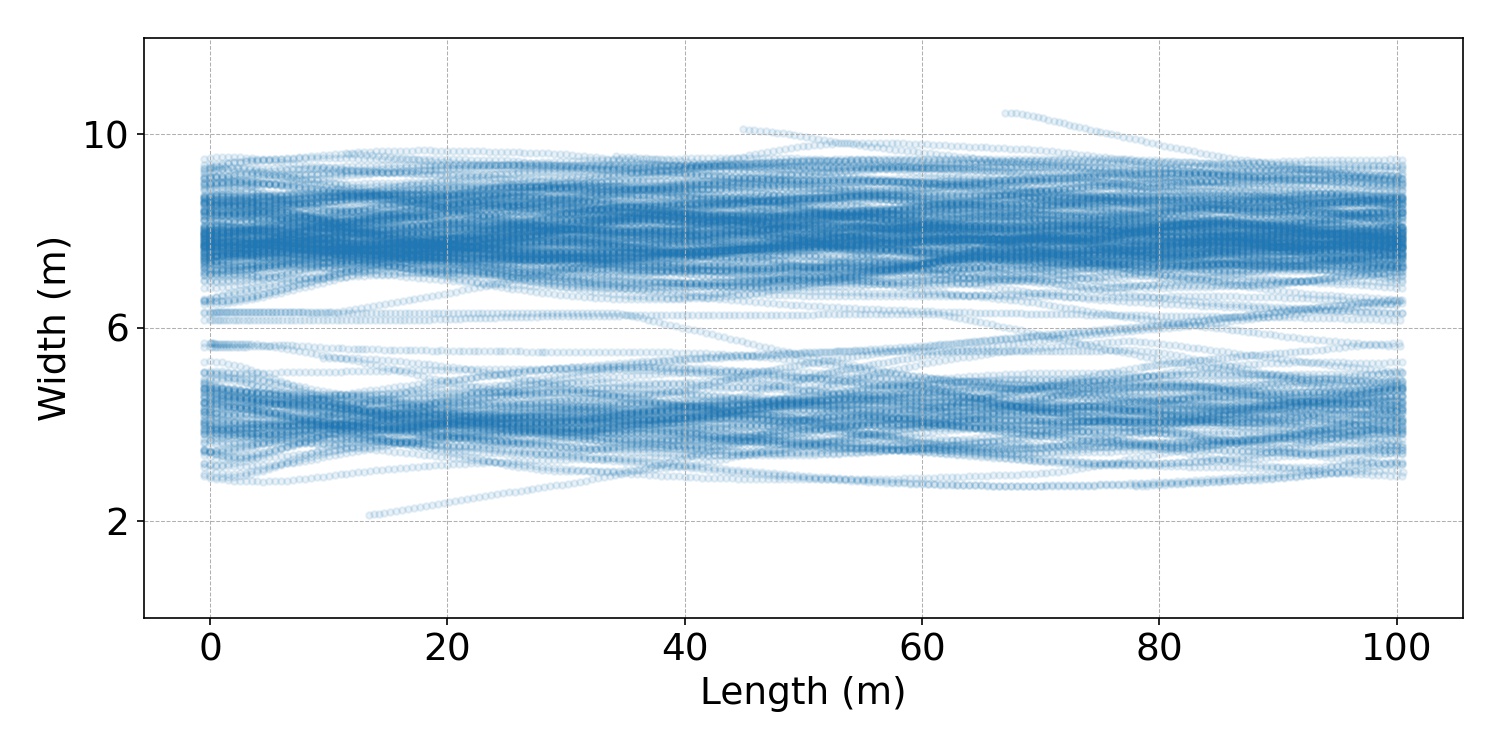}
    \caption{Fully formed virtual lanes, 5min simulation.}\label{subfig:virtual}
\end{subfigure}
\caption{Virtual lanes under congested traffic conditions are observed in the pNEUMA dataset and are also reproduced in simulations with the hybrid model. Motorcycles take advantage of their maneuverability and move in the available space between cars.}\label{fig:result}
\end{figure}

Figure~\ref{fig:result} corroborates the ability of the hybrid model to reproduce the phenomenon of virtual lanes faithfully. Under congested traffic conditions, motorcycles take advantage of their smaller size and better maneuverability and travel in the available space beteen the cars. Although this is a known phenomenon, its dynamics are not yet well understood and this is a first step towards a more detailed study of lane formation in mixed vehicular traffic. In contrast to second-order models, where inertial effects may lead to local gridlocks, the anticipation and collision-free properties of the hybrid model coupled with appropriate choice of parameters, lead to very rapid lane formation. It is worth mentioning that fully formed lanes can be observed already after a simulation time of 5min. During this time, which corresponds to 6000 integration steps, no collision was observed. The respective histograms of lateral positions from the simulated trajectories and a 15min sample from the pNEUMA data at a location quite far from a traffic light without significant queue lengths, show good agreement on a qualitative level.

In future research, it would be interesting to investigate the lane formation process itself. Do virtual lanes appear gradually or abruptly when a critical threshold is exceeded, similarly to phase-transitions? Experimentation with the hybrid model and few results from the literature \cite{Vlahogianni2014}, suggest that density and speed differences may be decisive factors. Towards the outlined direction, it would be beneficial to come up with some quantifiable lane formation index, as similar metrics have been proposed in pedestrian traffic theory \cite{Schadschneider2018}. This can be for example km travelled in virtual lanes. Regarding the realism of the hybrid model, we acknowledge that some idealized assumptions can be lifted in subsequent iterations. Clearly, traffic signals should be incorporated in a more comprehensive simulator. As far as the initialization is concerned, dart-throwing can be extended to handle simultaneous speed assignment with variable disk size (based on speed) in order to guarantee crash-free conditions. Furthermore, the movement of cars should be more realistic, allowing for lane-changes \cite{Kesting2007, Toledo2007, Talebpour2015}. Finally, the modular character of the hybrid model can facilitate the incorporation of more advanced car-following models. For a recent review on the state of the art in car-following dynamics, see \cite{Punzo2021}.

Of course lane formation is just one application of the hybrid model. There exists a body of literature on traffic flow instabilities, such as the formation and propagation of stop-and-go waves \cite{Yeo2009, Laval2010, Makridis2020}, but these works neglect the potentially destabilizing role of motorcycles. Last but not least, a better theoretical understanding of mixed traffic dynamics can produce significant results in more concrete applications such as traffic safety, especially taking into account questions raised by the proliferation of connected and automated vehicles.

\section{Conclusion}
In this article we introduce the concept of mode-dependent lane discipline as an alternative to the traditional distinction between homogeneous and heterogeneous traffic. Empirical evidence from drone videography, as exemplified by the massive pNEUMA dataset, shows that motorcycles participate in self-organized lane formation that is a result of complex modal interactions between different traffic participants competing for limited space, while cars simply adhere to agreed-upon social norms and follow the lanes as defined by the road infrastructure. In fact, the very notion of lane is revisited and the idea of generalized lane-keeping envelops is introduced. The latter also covers virtual lanes. One of the central findings of this paper is the existence of sparsity during the execution of lateral maneuvers that greatly simplifies the task of lane identification in a way that is robust to systematic bias. The result is a very detailed trajectory segmentation that reveals significant modal differences with respect to the spatial allocation of lanes. Finally, inspired by models from pedestrian dynamics, we propose a hybrid model with anticipatory and collision-free properties that is able to reproduce faithfully the empirically observed phenomena of self-organized lane formation.

\section{Acknowledgements}
The authors would like to thank Prof. Armin Seyfried for the inspiring discussions on the links between pedestrian and vehicular traffic. This research was partially funded by Swiss National Science Foundation (SNSF) grant \href{https://www.sciencedirect.com/science/article/pii/S0968090X19310320}{(200021\_188590)} “pNEUMA: On the new era of urban traffic models with massive empirical data from aerial footage”.

\newpage

\bibliographystyle{trb}
\bibliography{trb_template}
\end{document}